\documentclass{llncs}

\usepackage[pdftex, pagebackref=true, pdfpagelabels=true,
hypertexnames=true, plainpages=false, naturalnames=false,
bookmarks=true, bookmarksnumbered=true,
pdfpagemode=UseOutlines,]{hyperref}
\hypersetup{colorlinks, citecolor=[rgb]{0.0,0.0,0.5},
filecolor=[rgb]{0.6,0.0,0.0}, linkcolor=[rgb]{0.0,0.0,0.5},
urlcolor=[rgb]{0.0,0.0,0.5}}
\usepackage{url}            
\usepackage[numbers]{natbib}

\usepackage{placeins}
\usepackage{graphicx, subcaption} \graphicspath{ {Images/} }

\usepackage{tabularx}
\usepackage{booktabs}
\usepackage{multirow}
\usepackage{bbding}
\begin{document}

\title{Deep Dilated Convolutional Nets for the Automatic Segmentation of Retinal Vessels} 

\author{Ali Hatamizadeh\inst{1} \and Hamid Hosseini\inst{2} \and Zhengyuan Liu\inst{1} \and Steven D.~Schwartz\inst{2} \and Demetri  Terzopoulos\inst{1}
}

\institute{Computer Science Department, Henry Samueli School of Engineering
\and
Stein Eye Institute, David Geffen School of Medicine\\ University of California, Los Angeles, CA 90095, USA\\
}

\authorrunning{Hatamizadeh et al.}

\maketitle

\begin{abstract}
The reliable segmentation of retinal vasculature can provide the means
to diagnose and monitor the progression of a variety of diseases
affecting the blood vessel network, including diabetes and
hypertension. We leverage the power of convolutional neural networks
to devise a reliable and fully automated method that can accurately
detect, segment, and analyze retinal vessels. In particular, we
propose a novel, fully convolutional deep neural network with an
encoder-decoder architecture that employs dilated spatial pyramid
pooling with multiple dilation rates to recover the lost content in
the encoder and add multiscale contextual information to the decoder.
We also propose a simple yet effective way of quantifying and tracking the widths of retinal vessels through direct use of the
segmentation predictions. Unlike previous deep-learning-based
approaches to retinal vessel segmentation that mainly rely on
patch-wise analysis, our proposed method leverages a whole-image
approach during training and inference, resulting in more efficient
training and faster inference through the access of global content in
the image. We have tested our method on two publicly available
datasets, and our state-of-the-art results on both the DRIVE and
CHASE-DB1 datasets attest to the effectiveness of our approach.
\end{abstract}

\section{Introduction}

The retina and its vasculature are directly visible due to the
optically clear media of the human eye. As the only part of the
central nervous system that can be rapidly and non-invasively imaged
with a variety of modalities in the out-patient setting, the retina
provides a window into the human body, thus offering the opportunity
to assess changes associated with systemic diseases such as
hypertension, diabetes, and neurodegenerative disorders. The sequelae
of these conditions, specifically stroke, heart disease, and dementia
represent major causes of morbidity and mortality in the developed
world. To date, all classification schemes for retinal vascular
changes in these conditions, particularly in the early stages of
disease, have been based on qualitative changes based on human
assessment. We and others hypothesize that biomarkers of seriously
adverse health events exist in the quantitative assessment of retinal
vasculature changes associated with early, even asymptomatic,
diabetes, hypertension, or neurodegenerations. Specifically, high
blood pressure, for example, causes structural changes in the macro-
and micro-vasculature of vital organs throughout the body, including
the brain, heart, and kidney. The retinal vasculature is similarly
impacted but has the advantage of accessibility to multimodal imaging,
providing the opportunity to quantitatively assess prognosis, risk,
and response to treatment. Narrowing of retinal vessels has been
described as an early, classic sign of hypertension. However, this
early sign is difficult to use in everyday clinical practice, which
usually includes only non-quantitative, subjective visual assessment
of the retina by examination, photograph, or even angiography. An
automated, quantitative, reliable, reproducible tool that measures
changes in the retinal vasculature in response to disease and
intervention might augment and disrupt current evaluation and
treatment paradigms by allowing physicians to detect disease, predict
outcomes, and assess interventions much earlier in the course of
disease, thereby opening the potential for improved outcomes in major
unmet public health needs.

A critical step in tracking important structural changes of the
retinal vasculature is segmentation of the retinal vessels, as it
enables locating the veins and arteries and extracting relevant
information such as a profile of the width changes of the vessels.
Since the manual segmentation of vessels by clinicians is a
notoriously laborious and error-prone process, it is important to
establish fully automated and reliable segmentation methods that can
be leveraged for extracting the aforementioned information with
minimal supervision.

Since the advent of deep learning, Convolutional Neural Networks
(CNNs) have become popular due to their powerful, nonlinear feature
extraction capabilities in many computer vision related applications
\citep{hatamizadeh2018automatic, kachuee2018dynamic,
Xie_2018_CVPR,zhu2019anatomynet,akula2019visual}. Several researchers have applied CNNs
to the task of retinal vessel segmentation in fundus images. However,
most are patch-wise methods that ignore the global context in the
image and are usually inefficient during inference.
\citet{melinvsvcak2015retinal} employed a simple 10-layer CNN
architecture based on a patch-wise technique, but their results suffer
from low sensitivity in comparison to other techniques.
\citet{fu2016deepvessel} treated the problem of segmentation as a
boundary detection problem and combined a CNN with a conditional
random field to address the segmentation of retinal vessels, but their
method is slower than whole-image CNNs and is outperformed by a
number of other proposed methods in different metrics.
\citet{zhuang2018laddernet} proposed an architecture based on U-Net
\citep{Ronneberger2015}, which utilizes multiple-path networks that
leverages a path-wise formulation in segmenting retinal vessels.

\begin{figure}[t]
\centering
\subcaptionbox{}{\includegraphics[width=0.32\linewidth,height=0.32\linewidth]{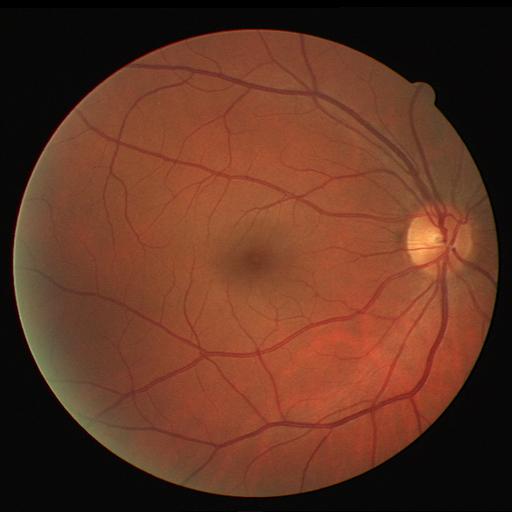}}
\hfill
\subcaptionbox{}{\includegraphics[width=0.32\linewidth,height=0.32\linewidth]{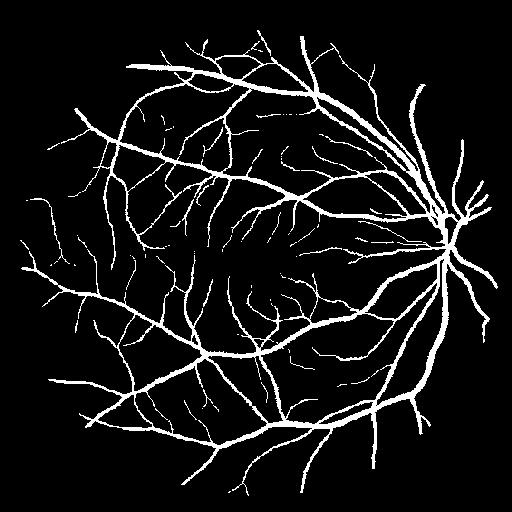}}
\hfill
\subcaptionbox{}{\includegraphics[width=0.32\linewidth,height=0.32\linewidth]{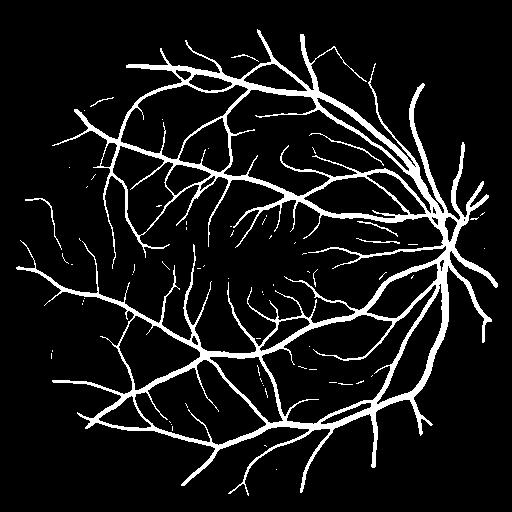}}
\caption{Segmentation of an image from the DRIVE dataset. (a) Input
image. (b) Ground truth. (c) Segmentation prediction output by our
network.}
\label{intro_fig}
\end{figure}

In the present paper, we exploit the power of CNNs to create a
reliable, fully automated, method that can accurately detect and
segment retinal vessels, and we devise an algorithm for the automatic
quantification of widths in retinal vessels directly from the
segmentation masks, which can be employed toward the creation the
aforementioned biomarkers. In particular, we introduce an
encoder-decoder CNN architecture that leverages a new dilated spatial
pyramid pooling with multiple dilation rates, which preserves
resolution yet adds multiscale information to the decoder.
Figure~\ref{intro_fig} shows an input fundus image with its
corresponding ground-truth and the vascular segmentation output by our
network.

\begin{figure*}[t]
\includegraphics[width=\textwidth]{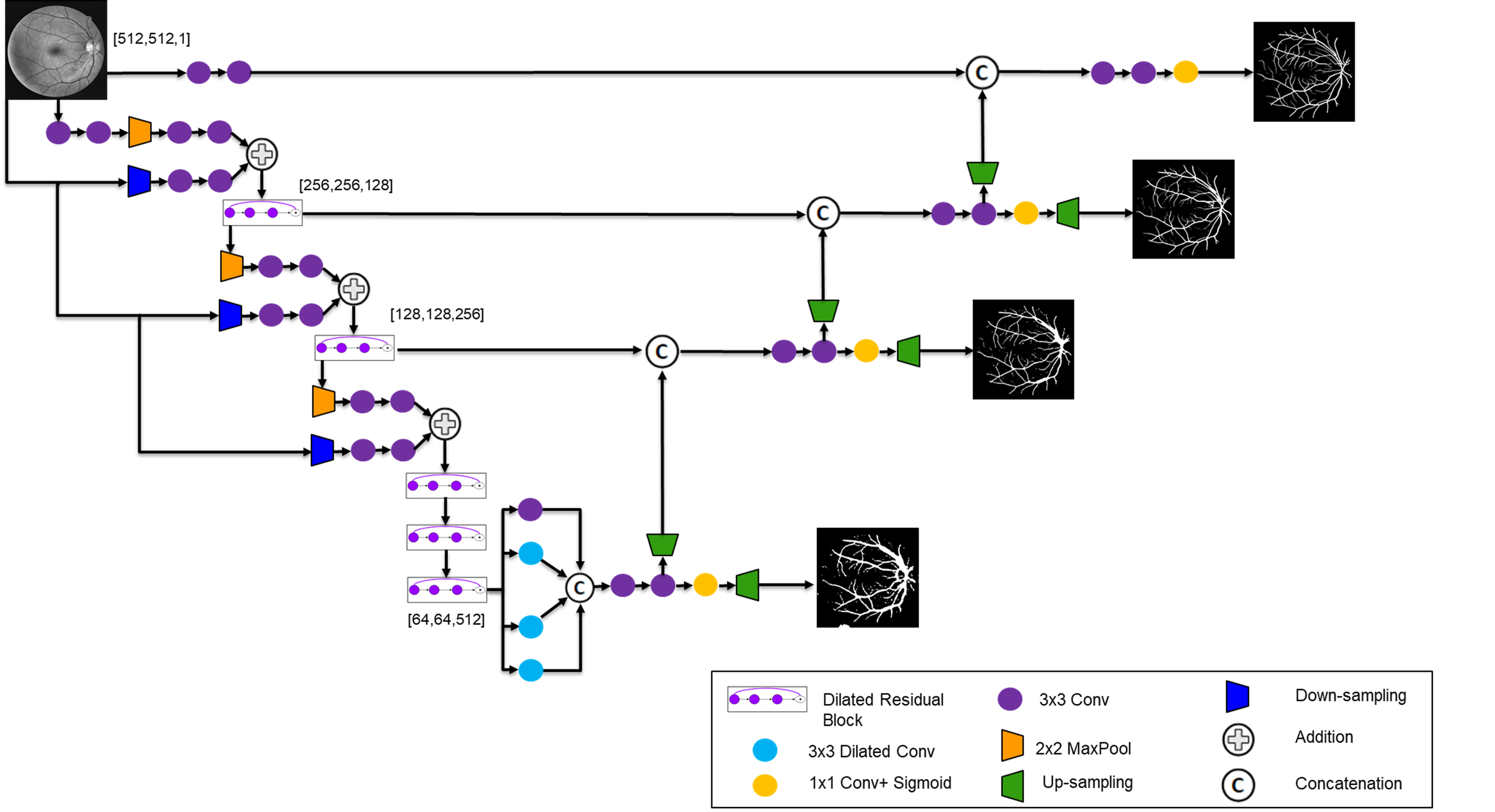}
\caption{Our proposed fully convolutional architecture. Dilated
spatial pyramid pooling aggregates the outputs of the multiple
stages.}
\label{fig:pipeline_vnet}
\end{figure*}

\section{Method}

\subsection{Vessel Segmentation}

We propose a fully convolutional encoder-decoder architecture, as
depicted in Figure~\ref{fig:pipeline_vnet}, which leverages dilated
residual blocks along with deep supervision at multiple scales for
effectively learning the multiresolution details of retinal vessels.
Each convolutional layer with kernel $W$ is followed by a rectified
linear unit (ReLU) $Re(X) = \max(0, X)$ and a batch normalization
$\textit{BN}_{\gamma,\beta}(X)$ with parameters $\gamma,\beta$ that
are learned during training. Consequently, every location $i$ in the
output of a convolutional layer followed by ReLU and batch
normalization can be represented as
\begin{equation}
Y(i) = \textit{BN}_{\gamma,\beta}(Re(\sum_{j=1}^{} X[i+j\cdot r]W[j]))),
\label{eq:conv}
\end{equation}
where $r$ is the dilation rate. We employ both standard and dilated
convolutional layers for which the value of $r$ in the former is $1$
and in the latter depends on where it is used. In this work, we utilize dilated residual blocks that consist of two consecutive dilated convolutional layers whose outputs are fused with the input.

Our encoder-decoder architecture spans four different resolutions. In
the encoder, each path consist of 2 consecutive $3 \times 3$
convolutional layers, followed by a dilated residual unit with a
dilation rate of 2. Before being fed into the dilated residual unit,
the output of these convolutional layers are added with the output
feature maps of another 2 consecutive $3 \times 3$ convolutional
layers that learn additional multiscale information from the re-sized
input image in that resolution. At the third stage of our
architecture, we utilize a series of 3 consecutive dilated residual
blocks with dilation rates of 1, 2, and 4, respectively. Finally, we
incorporate a dilated spatial pyramid pooling layer with 4 different
dilation rates of 1, 6, 12 and 18 in order to recover the content lost
in the learned feature maps during the encoding process.

Subsequently, the decoder in our architecture receives the learned
multiscale contextual information of the dilated spatial pyramid
pooling layer and is connected to the dilated residual units at each
resolution via skip connections. In each path of the decoder, the
image is up-sampled and 2 consecutive $3 \times 3$ convolutional
layers are used before proceeding to the next resolution. Moreover,
each scale branches to an additional convolutional layer whose output
is resized to the original input image size and is followed by another
convolutional layer with sigmoid activation function.

These multiscale prediction maps contribute to the final loss layer.
We utilize a soft Sørensen-Dice loss function as our basis and
aggregate throughout each of the four resolutions:
\begin{equation}
\textit{Loss}=\sum_{m=1}^{4}(1- \sum_{n=1}^{N} {\frac{2G_{n}
P_{n,m}}{G_{n}+P_{n,m}+\epsilon}})+\lambda \|w \|_{2}^{2},
\label{eq:10th}
\end{equation} 
where $N$, $P_{n,m}$, and $G_{n}$ denote the total number of pixels,
the label prediction of pixel $n$ in scale $m$, and the ground truth
label of pixel $n$, respectively, $\epsilon$ is a smoothing constant,
and $\lambda$ is the weight decay regularization hyper-parameter.

\begin{figure}[t]
\centering
\includegraphics[width=\textwidth,height=\textheight,keepaspectratio]{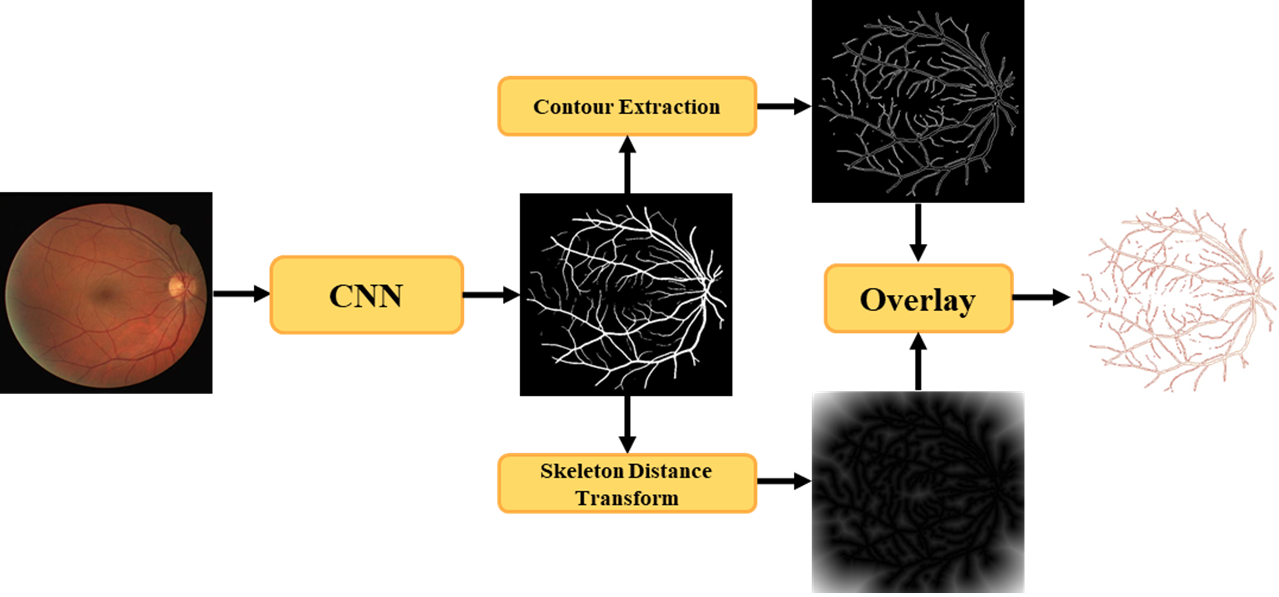}
\caption{Estimating the width profile of retinal vessels from
segmentation.}
\label{fig_post}
\end{figure}

\subsection{Vessel Width Estimation}

We propose a simple, yet effective method for the automatic estimation
of vessel width profiles by leveraging the segmentation masks obtained
by our CNN. Similar to \citep{Zhang:1984:FPA:357994.358023}, we first
obtain the skeleton of the image by successively identifying the
borderline pixels and removing the corresponding pixels that maintain
the connectivity of the vessels. This operation approximates the
center line of the vessel and represents its topology. We then
calculate the distance of each pixel to the derived center-lines by
applying an Euclidean distance transform to the generated feature map.
Finally, we extract the contour of the original segmentation mask and
overlay the generated distance transform onto this map to create the
final width map of the retinal vessels. Needless to say, our
formulation is only valid in areas where these vessels exist,
otherwise the width value is set to zero. Figure~\ref{fig_post}
illustrates our width estimation algorithm in more detail. Unlike
competing methods, our method does not rely on hand-crafted geometric
equations nor on user interaction.

\section{Experiments}

\subsection{Implementation Details}

We have implemented our CNN in TensorFlow \citep{abadi2016tensorflow}.
All the input images are converted to gray-scale, transformed by
contrast-limited adaptive histogram equalization, resized to a
predefined size of $512\times512$, and intensity normalized between 0
and 1. Our model is trained, with a batch size of 2, on an Nvidia
Titan XP GPU and an Intel® Core™ i7-7700K CPU @ 4.20GHz. We use the
Adam optimization algorithm with an initial learning rate of 0.001 and
exponentially decay its rate. The smoothing constant in the loss
function and the weight decay hyper-parameter are set to $10^{-5}$ and
$0.0008$, respectively. Since the number of images is limited, we
perform common data augmentation techniques such as rotating, flipping
horizontally and vertically, and transposing the image.

\subsection{Datasets}

We have tested our model on two publicly available retinal vessel
segmentation datasets---DRIVE and CHASE-DB1. The DRIVE dataset
consists of 40 two-dimensional RGB images with each image having a
resolution of $565 \times 584$ pixels, divided into a training set and
test set, each comprising 20 images. The CHASE-DB1 dataset includes 28
images, collected from both eyes of 14 children. Each image has a
resolution of $999 \times 960$ pixels. We divided the CHASE-DB1
dataset into a training set of 20 images and a testing set of 8
images.

\subsection{Results}

\begin{table}[t]
\centering
\caption{Segmentation Evaluations on the DRIVE and CHASE-DB1 datasets.}\label{tab_res}
\setlength{\tabcolsep}{0.7mm}
\begin{tabular}{r|c|c|c|c|c|c|c|c}
\hline
Method & \multicolumn{4}{|c|}{DRIVE} & \multicolumn{4}{|c}{CHASE-DB1}\\
\cline{2-9}
 & SE & SP & Acc & F1 & SE & SP & Acc & F1\\
\hline
\citet{melinvsvcak2015retinal} & 0.7276 & 0.9785 & 0.9466 & - & - & - & - & -\\
\citet{li2016cross} & 0.7569 & 0.9816 & 0.9527 & - & 0.7507 & 0.9793 & 0.9581 & -\\
Liskowski et al.~\citep{liskowski2016segmenting} & 0.7520 & 0.9806 & 0.9515 & - & - & - & - & -\\
\citet{fu2016deepvessel} & 0.7603 & - & 0.9523 & - & 0.7130 & - & 0.9489 & -\\
\citet{oliveira2018retinal} & 0.8039 & 0.9804 & 0.9576 & - & 0.7779 & \textbf{0.9864} & 0.9653 & -\\
M2U-Net \citep{laibacher2018m2u} & - & - & 0.9630 & 0.8091 & - & - & 0.9703 & 0.8006\\
U-Net \citep{alom2018recurrent} & 0.7537 & 0.9820 & 0.9531 & 0.8142 & 0.8288 & 0.9701 & 0.9578 & 0.7783\\
Recurrent U-Net \citep{alom2018recurrent} & 0.7751 & 0.9816 & 0.9556 & 0.8155 & 0.7459 & 0.9836  & 0.9622 & 0.7810\\
R2U-Net \citep{alom2018recurrent} & 0.7792 & 0.9816 & 0.9556 & 0.8171 & 0.7756 & 0.9820 & 0.9634 & 0.7928\\
LadderNet \citep{zhuang2018laddernet} & 0.7856 & 0.9810 & 0.9561 & 0.8202 & 0.7978 & 0.9818 & 0.9656 & 0.8031\\
DUNet \citep{jin2019dunet} & 0.7894  & \textbf{0.9870} & \textbf{0.9697} & 0.8203 & 0.8229 & 0.9821 & 0.9724 & 0.7853\\

\textbf{Ours} & \textbf{0.8197} & 0.9819 & 0.9686 & \textbf{0.8223} & \textbf{0.8300} & 0.9848 & \textbf{0.9750} & \textbf{0.8073}\\
\hline
\end{tabular}
\end{table}

\def\TP{\textit{TP}}
\def\TN{\textit{TN}}
\def\FP{\textit{FP}}
\def\FN{\textit{FN}}

We used the following metrics to measure the performance of our model:
With $\TP$, $\TN$, $\FP$, $\FN$ denoting true positive, true negative,
false positive, and false negative, respectively, the sensitivity and
specificity are given as
\begin{equation}
\textit{SE} = \frac{\TP}{\TP+\FN}, \qquad \textit{SP} =
\frac{\TN}{\TN+\FP},
\end{equation}
the accuracy and precision as
\begin{equation}
\textit{Acc} = \frac{\TP+\TN}{\TP+\TN+\FP+\FN}, \qquad
\textit{Precision} = \frac{\TP}{\TP+\FP},
\end{equation}
and the recall as
\begin{equation}
\textit{Recall} = \frac{\TP}{\TP+\FN}.
\end{equation}
The score
\begin{equation}
\textit{F1} = 2 \times \frac{\textit{Precision} \times
\textit{Recall}}{\textit{Precision} + \textit{Recall}}
\end{equation}
is the same as the Dice coefficient.

\begin{figure}[t]
\centering
\includegraphics[width=0.48\linewidth,height=0.48\linewidth,keepaspectratio]{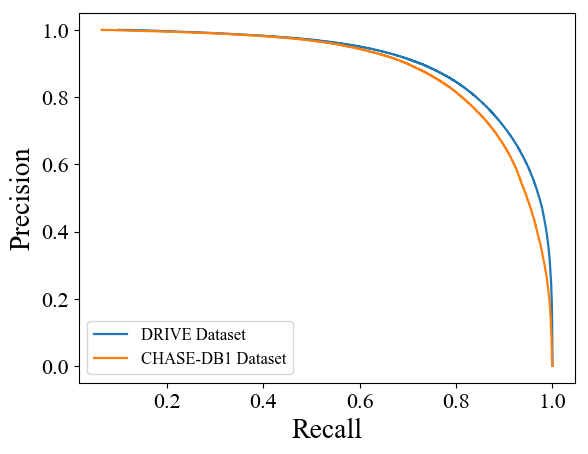}
\hfill
\includegraphics[width=0.48\linewidth,height=0.372\linewidth]{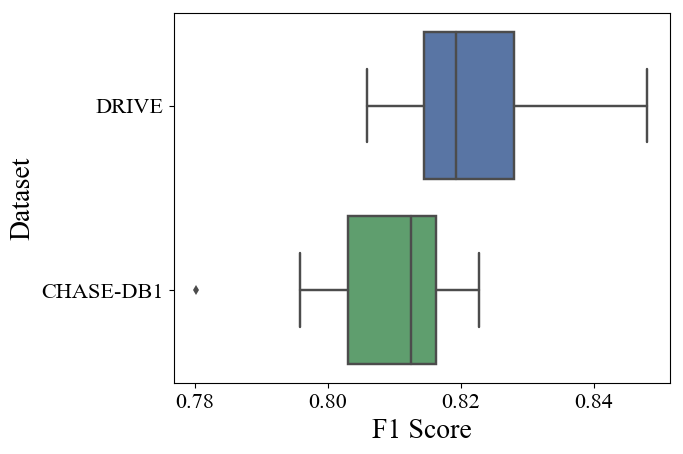}\\[4pt]
\caption{Evaluation of our model's performance on the DRIVE and CHASE-DB1 datasets.}
\label{roc}
\end{figure}

Table~\ref{tab_res} compares to competing methods the performance of
our model on the DRIVE and CHASE-DB1 datasets. On the DRIVE dataset,
our model exceeds the state-of-the-art performance on the F1 score and
sensitivity while being very competitive to \cite{jin2019dunet} in
specificity and accuracy. Figure~\ref{roc} presents the
precision-recall curve as well as the box and whisker plot of our
model's performance on the DRIVE dataset. The F1 scores are between
0.80 to 0.84 with no outliers falling outside the interquartile range.

In addition, for the CHASE-DB1 dataset our model exceeds the
state-of-the-art performance for all metrics except specificity, for
which \cite{oliveira2018retinal} performs slightly better. The F1
scores are between 0.78 to 0.82. The precision-recall curve in
Figure~\ref{roc} demonstrates a similar performance on both th DRIVE
and CHASE-DB1 datasets with a marginal lead for the latter dataset.

\begin{figure}
\centering
\vspace{-10pt}
\includegraphics[width=0.24\linewidth,height=0.24\linewidth]{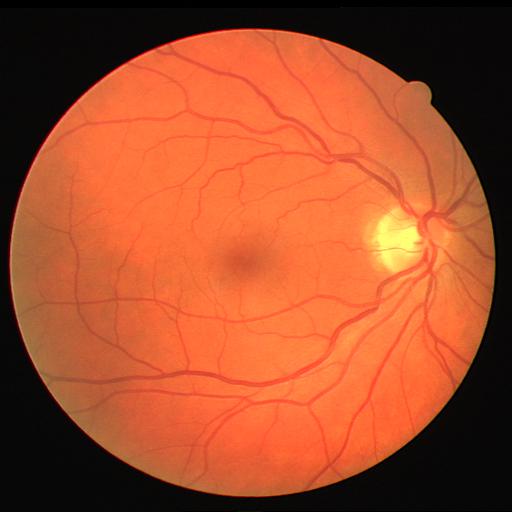}
\hfill
\includegraphics[width=0.24\linewidth,height=0.24\linewidth]{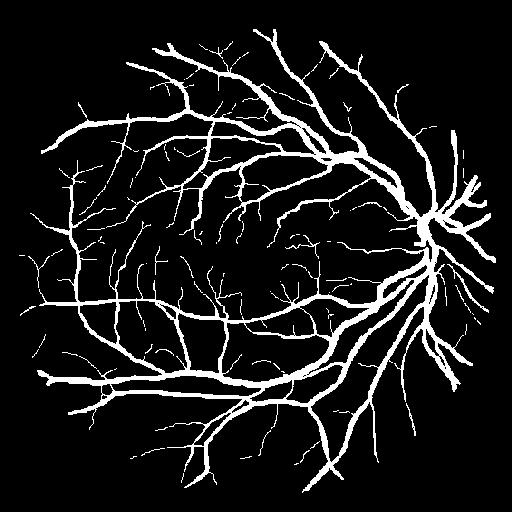}
\hfill
\includegraphics[width=0.24\linewidth,height=0.24\linewidth]{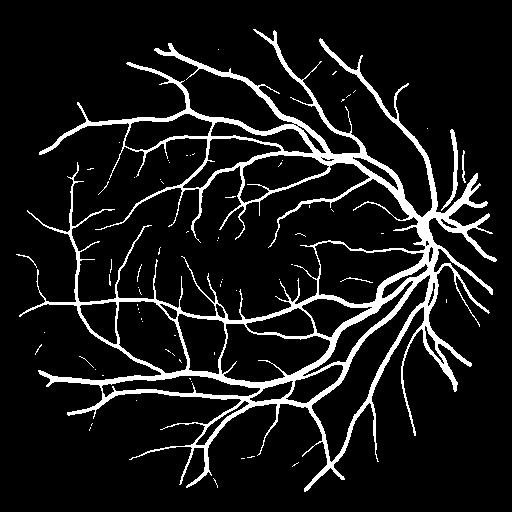}
\hfill
\includegraphics[width=0.24\linewidth,height=0.24\linewidth]{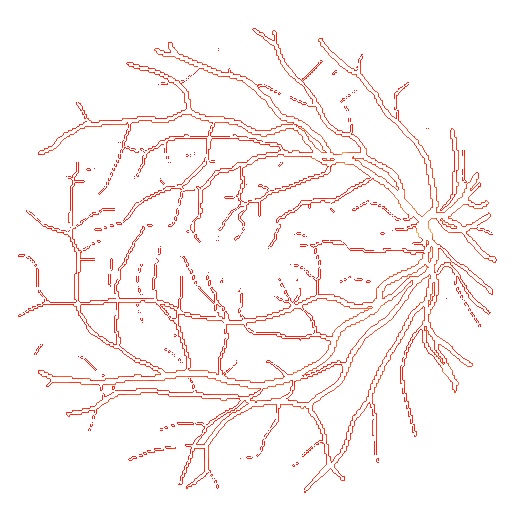}\\[4pt]

\includegraphics[width=0.24\linewidth,height=0.24\linewidth]{19_.jpg}
\hfill
\includegraphics[width=0.24\linewidth,height=0.24\linewidth]{19_label.jpg}
\hfill
\includegraphics[width=0.24\linewidth,height=0.24\linewidth]{19_predimg.jpg}
\hfill
\includegraphics[width=0.24\linewidth,height=0.24\linewidth]{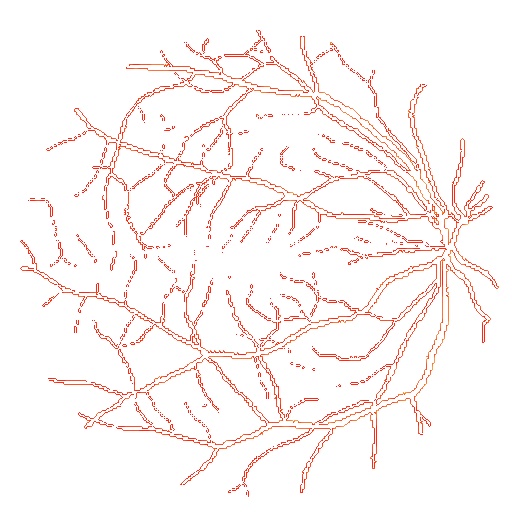}\\[4pt]

\includegraphics[width=0.24\linewidth,height=0.24\linewidth]{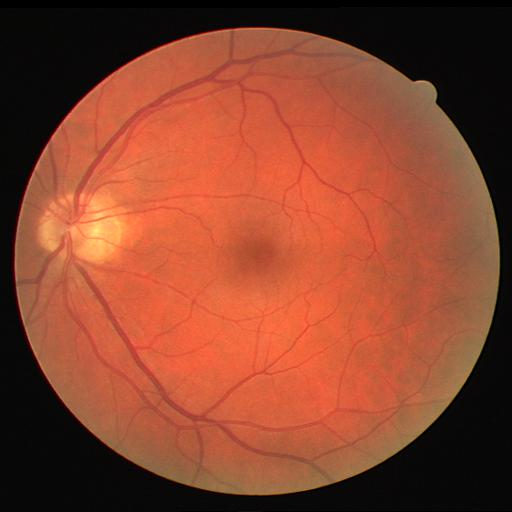}
\hfill
\includegraphics[width=0.24\linewidth,height=0.24\linewidth]{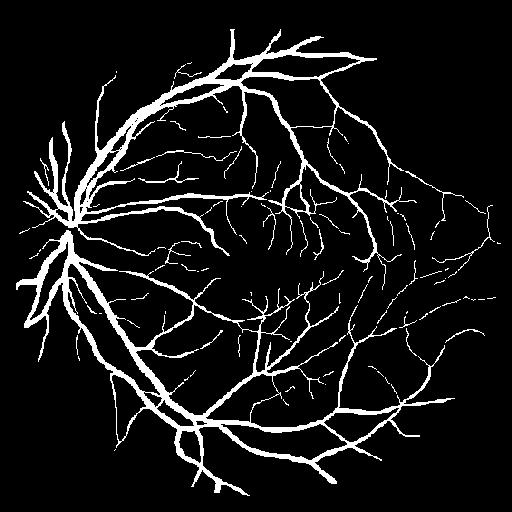}
\hfill
\includegraphics[width=0.24\linewidth,height=0.24\linewidth]{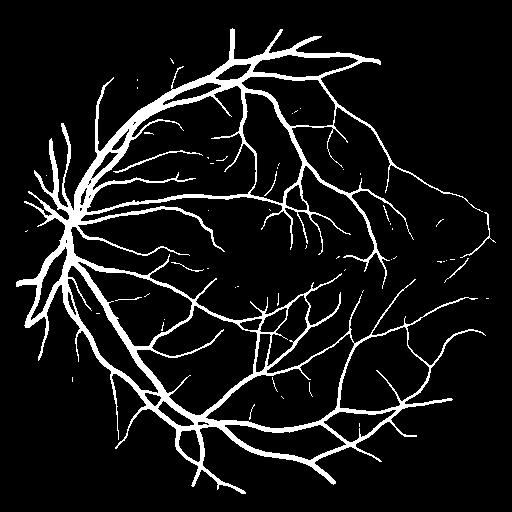}
\hfill
\includegraphics[width=0.24\linewidth,height=0.24\linewidth]{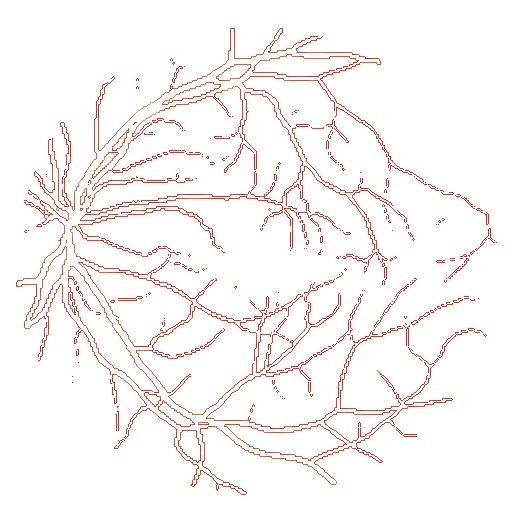}\\[4pt]

\includegraphics[width=0.24\linewidth,height=0.24\linewidth]{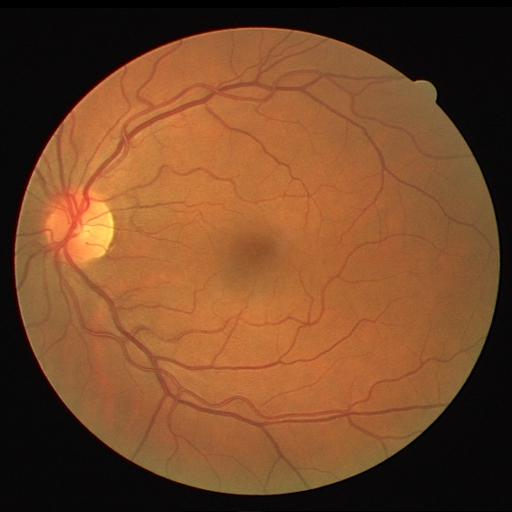}
\hfill
\includegraphics[width=0.24\linewidth,height=0.24\linewidth]{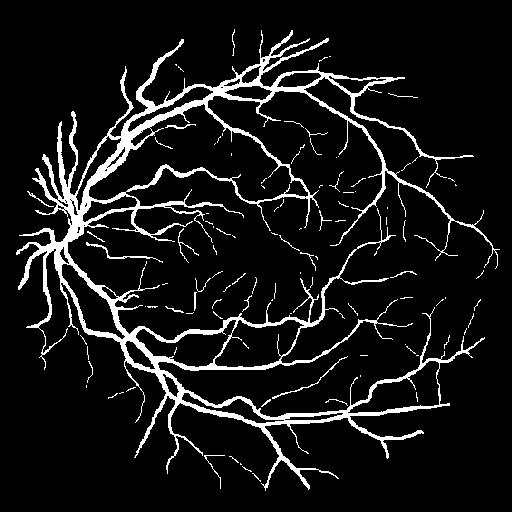}
\hfill
\includegraphics[width=0.24\linewidth,height=0.24\linewidth]{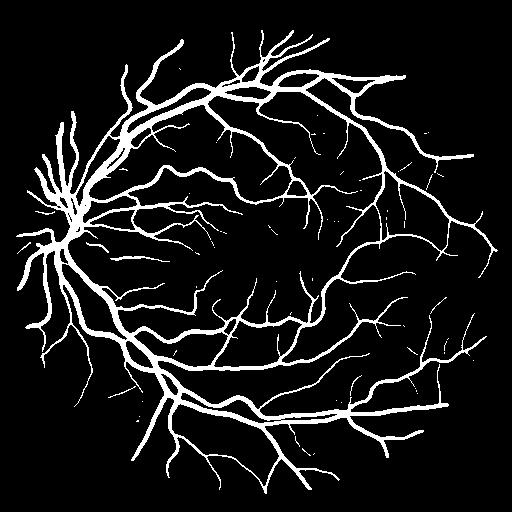}
\hfill
\includegraphics[width=0.24\linewidth,height=0.24\linewidth]{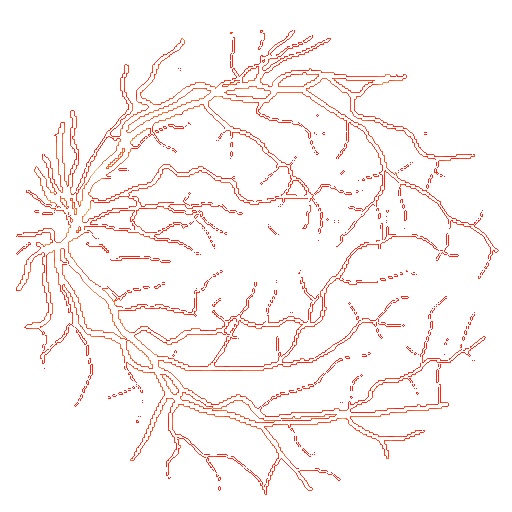}\\[4pt]

\includegraphics[width=0.24\linewidth,height=0.24\linewidth]{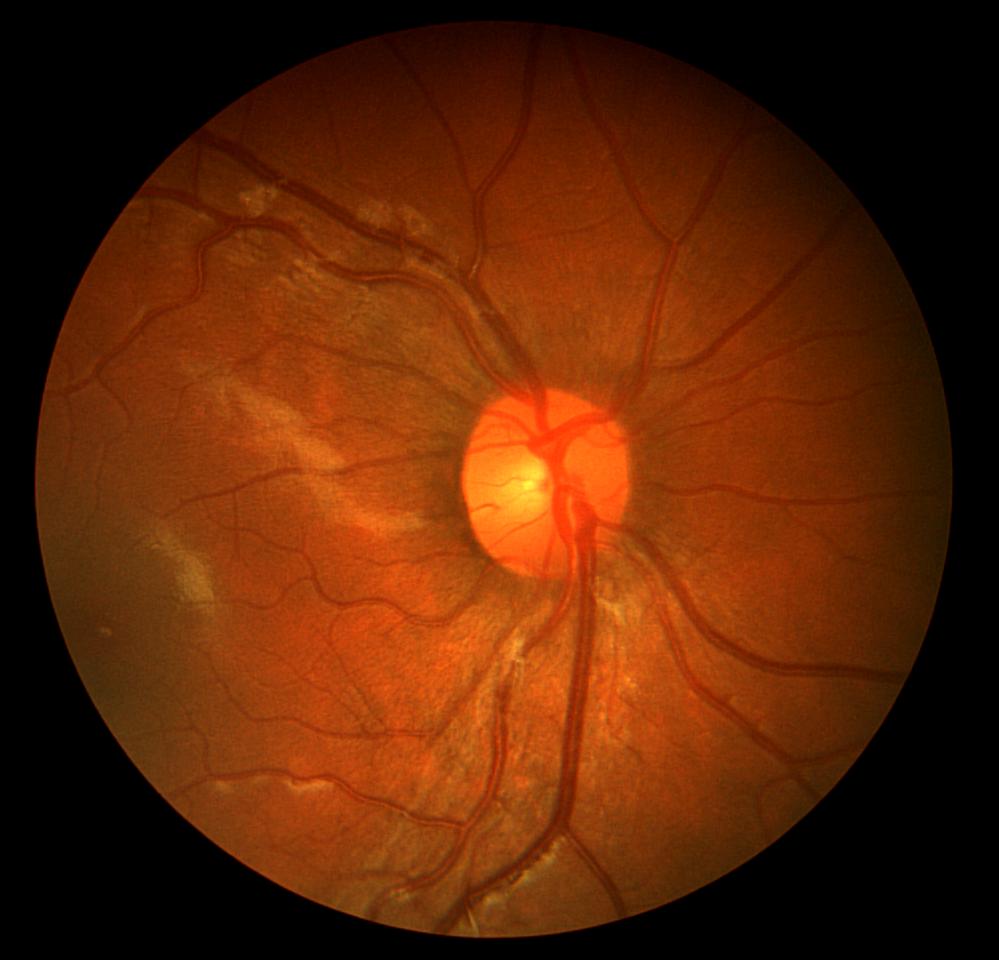}
\hfill
\includegraphics[width=0.24\linewidth,height=0.24\linewidth]{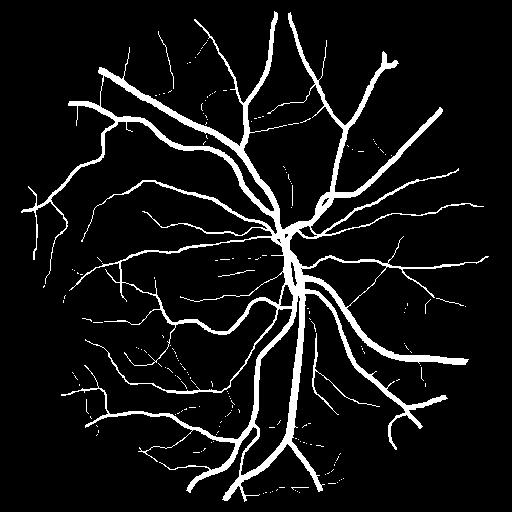}
\hfill
\includegraphics[width=0.24\linewidth,height=0.24\linewidth]{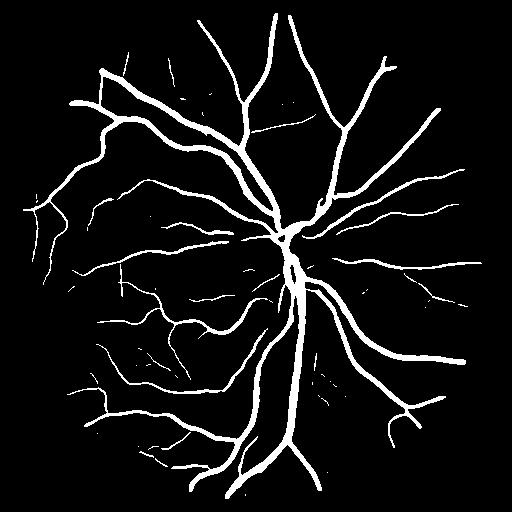}
\hfill
\includegraphics[width=0.24\linewidth,height=0.24\linewidth]{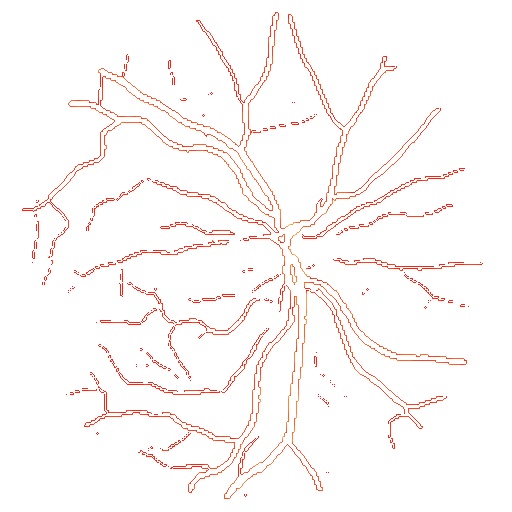}\\[4pt]

\includegraphics[width=0.24\linewidth,height=0.24\linewidth]{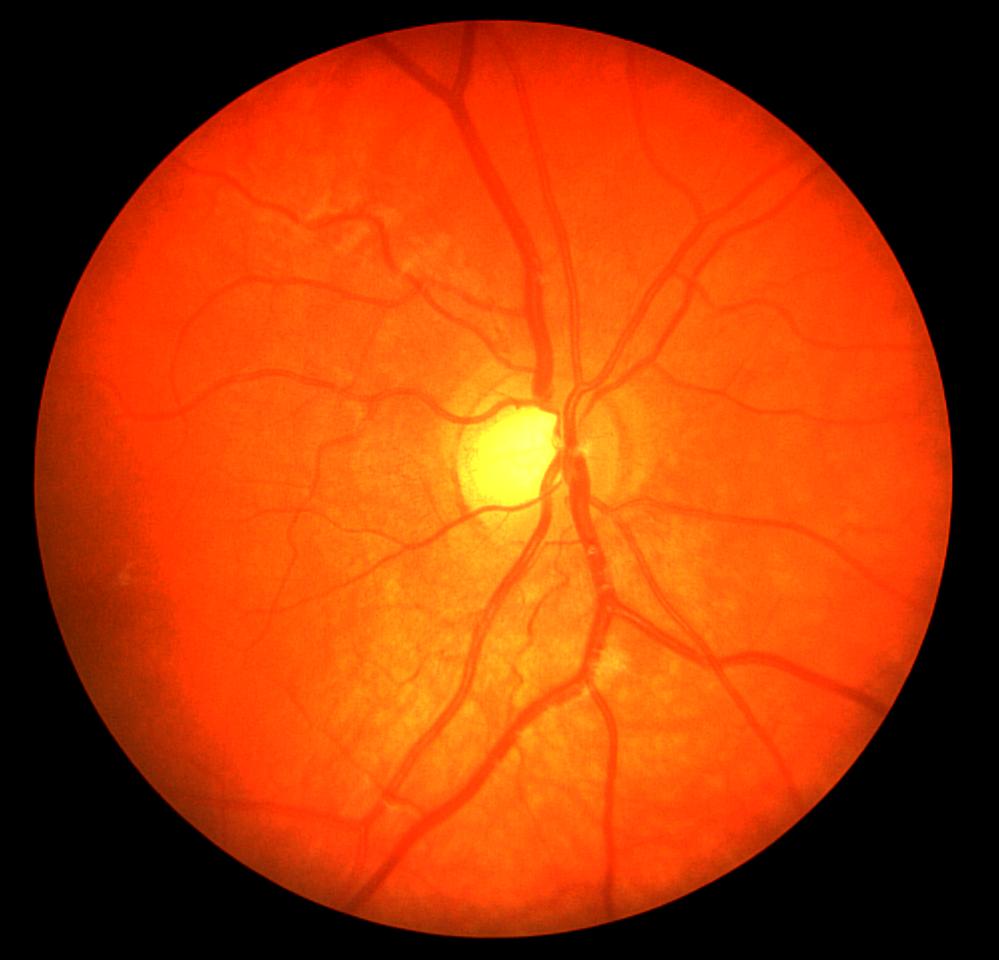}
\hfill
\includegraphics[width=0.24\linewidth,height=0.24\linewidth]{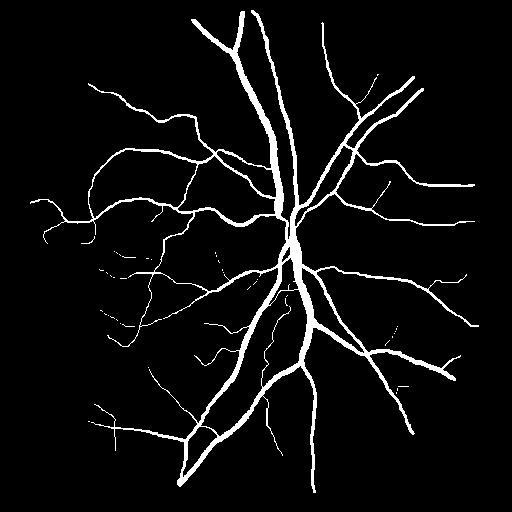}
\hfill
\includegraphics[width=0.24\linewidth,height=0.24\linewidth]{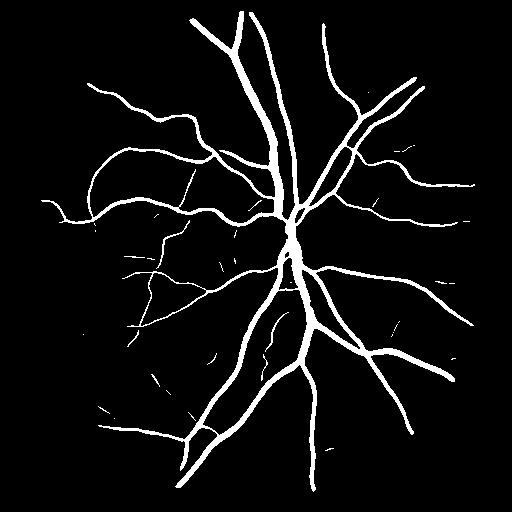}
\hfill
\includegraphics[width=0.24\linewidth,height=0.24\linewidth]{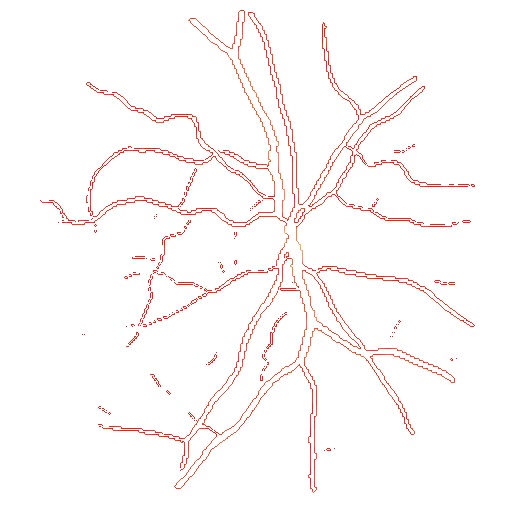}\\[4pt]

\makebox[0.24\linewidth]{(a)} \hfill \makebox[0.24\linewidth]{(b)}
\hfill \makebox[0.24\linewidth]{(c)} \hfill
\makebox[0.24\linewidth]{(d)}

\caption{(a) Input images (rows 1--4 from DRIVE; rows 5--6 from
CHASE-DB1), (b) labels, (c) segmentations, (d) width estimation
profiles.}
\label{fig_data}
\end{figure}

For 4 images from the DRIVE dataset and 2 images from the CHASE-DB1
dataset, Figure~\ref{fig_data} demonstrates the final segmentation of
our model compared to the masks that were manually created by
clinicians as well as the generated width profiles. Clearly, our model
successfully captures the intricate arteries and veins without the
presence of any additional false positives as are present in the
outputs of the competing methods.

\section{Discussion}

Our substantially improved state-of-the-art results on two publicly
available datasets, DRIVE and CHASE-DB1, confirm the effectiveness of
our model. Unlike the competing patch-wise approaches, our method
operates on the entire image. This has proven to be beneficial as our
model is significantly faster than the patch-wise approaches that must
slide moving windows of multiple sizes over the image. Additionally,
our model produces more natural and continuous segmentation masks and
is able to capture finer details because it benefits from a dilated
spatial pyramid pooling layer that recovers the content lost during
encoding by leveraging different dilation rates and aggregating the
multiscale feature maps into the decoder. Furthermore, introducing the
input image at multiple scales throughout our architecture and
introducing supervision at each of these scales helps our model to
effectively aggregate the outputs of different stages.

Our technique for estimating the width profiles of retinal vessels by
leveraging the generated segmentation masks has also proven to be
effective. Its accuracy promises to help in quantifying new relevant
biomarkers that correlate with the narrowing and structural changes of
vessels.

\section{Conclusion}

We have presented a novel, fully automated method for retinal vessel
segmentation and width estimation. Our deep CNN employs spatial
dilated pyramid pooling and introduces the input image at multiple
scales with supervision to segment retinal vessels in order to capture
the smallest structural details. Our method was tested on two publicly
available datasets. It has achieved better than state-of-the-art
results in sensitivity and accuracy while being comparable in F1 score
on the DRIVE dataset. It also achieves competitive results on the
CHASE-DB1 dataset. In addition, we have introduced a method that
employs the vessel segmentation maps to estimate the width profiles of
retinal vessels. Such information may be very helpful to clinicians as
they explore novel biomarkers and in the quantitative assessment of
retinal vasculature changes associated with diseases such as diabetes
and hypertension.

\bibliography{paper92}

\begin{thebibliography}{17}
\providecommand{\natexlab}[1]{#1}
\providecommand{\url}[1]{\texttt{#1}}
\providecommand{\urlprefix}{}

\bibitem[{Abadi et~al.(2016)Abadi, Barham, Chen, Chen, Davis, Dean, Devin,
  Ghemawat, Irving, Isard et~al.}]{abadi2016tensorflow}
Abadi, M., Barham, P., Chen, J., Chen, Z., Davis, A., Dean, J., Devin, M.,
  Ghemawat, S., Irving, G., Isard, M., et~al.: Tensorflow: {A} system for
  large-scale machine learning.
\newblock In: OSDI. vol.~16, pp. 265--283 (2016)

\bibitem[{Akula and Zhu(2019)}]{akula2019visual}
Akula, A.R., Zhu, S.C.: Visual discourse parsing.
\newblock arXiv preprint arXiv:1903.02252  (2019)

\bibitem[{Alom et~al.(2018)Alom, Hasan, Yakopcic, Taha, and
  Asari}]{alom2018recurrent}
Alom, M.Z., Hasan, M., Yakopcic, C., Taha, T.M., Asari, V.K.: Recurrent
  residual convolutional neural network based on {U-Net} ({R2U-Net}) for
  medical image segmentation.
\newblock arXiv preprint arXiv:1802.06955  (2018)

\bibitem[{Fu et~al.(2016)Fu, Xu, Lin, Wong, and Liu}]{fu2016deepvessel}
Fu, H., Xu, Y., Lin, S., Wong, D.W.K., Liu, J.: Deepvessel: Retinal vessel
  segmentation via deep learning and conditional random field.
\newblock In: International Conference on Medical Image Computing and
  Computer-Assisted Intervention. pp. 132--139. Springer (2016)

\bibitem[{Imran et~al.(2018)Imran, Hatamizadeh, Ananth, Ding, Terzopoulos, and
  Tajbakhsh}]{hatamizadeh2018automatic}
Imran, A., Hatamizadeh, A., Ananth, S.P., Ding, X., Terzopoulos, D., Tajbakhsh,
  N.: Automatic segmentation of pulmonary lobes using a progressive dense
  {V}-network.
\newblock In: Deep Learning in Medical Image Analysis and Multimodal Learning
  for Clinical Decision Support, pp. 282--290. Springer (2018), {Proc.} Fourth
  MICCAI International Workshop on Deep Learning in Medical Image Analysis
  (DLMIA 18)

\bibitem[{Jin et~al.(2019)Jin, Meng, Pham, Chen, Wei, and Su}]{jin2019dunet}
Jin, Q., Meng, Z., Pham, T.D., Chen, Q., Wei, L., Su, R.: Dunet: A deformable
  network for retinal vessel segmentation.
\newblock Knowledge-Based Systems  (2019)

\bibitem[{Kachuee et~al.(2018)Kachuee, Darabi, Moatamed, and
  Sarrafzadeh}]{kachuee2018dynamic}
Kachuee, M., Darabi, S., Moatamed, B., Sarrafzadeh, M.: Dynamic feature
  acquisition using denoising autoencoders.
\newblock IEEE Transactions on Neural Networks and Learning Systems pp. 1--11
  (2018)

\bibitem[{Laibacher et~al.(2018)Laibacher, Weyde, and
  Jalali}]{laibacher2018m2u}
Laibacher, T., Weyde, T., Jalali, S.: {M2U-Net}: Effective and efficient
  retinal vessel segmentation for resource-constrained environments.
\newblock arXiv preprint arXiv:1811.07738  (2018)

\bibitem[{Li et~al.(2016)Li, Feng, Xie, Liang, Zhang, and Wang}]{li2016cross}
Li, Q., Feng, B., Xie, L., Liang, P., Zhang, H., Wang, T.: A cross-modality
  learning approach for vessel segmentation in retinal images.
\newblock IEEE Transactions on Medical Imaging 35(1), 109--118 (2016)

\bibitem[{Liskowski and Krawiec(2016)}]{liskowski2016segmenting}
Liskowski, P., Krawiec, K.: Segmenting retinal blood vessels with deep neural
  networks.
\newblock IEEE Transactions on Medical Imaging 35(11), 2369--2380 (2016)

\bibitem[{Melin{\v{s}}{\v{c}}ak et~al.(2015)Melin{\v{s}}{\v{c}}ak,
  Prenta{\v{s}}i{\'c}, and Lon{\v{c}}ari{\'c}}]{melinvsvcak2015retinal}
Melin{\v{s}}{\v{c}}ak, M., Prenta{\v{s}}i{\'c}, P., Lon{\v{c}}ari{\'c}, S.:
  Retinal vessel segmentation using deep neural networks.
\newblock In: VISAPP 2015 (10th International Conference on Computer Vision
  Theory and Applications) (2015)

\bibitem[{Oliveira et~al.(2018)Oliveira, Pereira, and
  Silva}]{oliveira2018retinal}
Oliveira, A.F.M., Pereira, S.R.M., Silva, C.A.B.: Retinal vessel segmentation
  based on fully convolutional neural networks.
\newblock Expert Systems with Applications  (2018)

\bibitem[{Ronneberger et~al.(2015)Ronneberger, Fischer, and
  Brox}]{Ronneberger2015}
Ronneberger, O., Fischer, P., Brox, T.: U-net: Convolutional networks for
  biomedical image segmentation.
\newblock In: Proc. of MICCAI. pp. 234--241. Springer (2015)

\bibitem[{Xie et~al.(2018)Xie, Zheng, Gao, Wang, Zhu, and
  Nian~Wu}]{Xie_2018_CVPR}
Xie, J., Zheng, Z., Gao, R., Wang, W., Zhu, S.C., Nian~Wu, Y.: Learning
  descriptor networks for {3D} shape synthesis and analysis.
\newblock In: IEEE Conference on Computer Vision and Pattern Recognition (CVPR)
  (June 2018)

\bibitem[{Zhang and Suen(1984)}]{Zhang:1984:FPA:357994.358023}
Zhang, T.Y., Suen, C.Y.: A fast parallel algorithm for thinning digital
  patterns.
\newblock Communications of the ACM 27(3), 236--239 (Mar 1984)

\bibitem[{Zhu et~al.(2019)Zhu, Huang, Zeng, Chen, Liu, Qian, Du, Fan, and
  Xie}]{zhu2019anatomynet}
Zhu, W., Huang, Y., Zeng, L., Chen, X., Liu, Y., Qian, Z., Du, N., Fan, W.,
  Xie, X.: Anatomynet: Deep learning for fast and fully automated whole-volume
  segmentation of head and neck anatomy.
\newblock Medical physics 46(2), 576--589 (2019)

\bibitem[{Zhuang(2018)}]{zhuang2018laddernet}
Zhuang, J.: {LadderNet}: Multi-path networks based on {U-Net} for medical image
  segmentation.
\newblock arXiv preprint arXiv:1810.07810  (2018)

\end{thebibliography}

\end{document}